\documentclass[aps,prb,reprint,superscriptaddress]{revtex4-1}
\usepackage{braket}
\usepackage{graphicx}
\usepackage{amsmath}
\usepackage[usenames]{color}

\begin{document}

%\preprint{}

\title{Extrinsic Spin Hall Effect from  Anisotropic Rashba Spin-Orbit Coupling in Graphene}
\author{H.-Y. Yang}
\affiliation{Department of Physics, National Tsing Hua University, Hsinchu 30013, Taiwan}
\author{Chunli Huang}
\affiliation{Division of Physics and Applied Physics, School of Physical and Mathematical Sciences, Nanyang
Technological University, Singapore 637371, Singapore}
\affiliation{Department of Physics, National Tsing Hua University, Hsinchu 30013, Taiwan}
\author{H. Ochoa} 
\affiliation{Fundaci\'on IMDEA Nanociencia, Cantoblanco 28049, Madrid, Spain.}
\author{M.~A. Cazalilla}
\affiliation{Department of Physics, National Tsing Hua University, Hsinchu 30013, Taiwan}
\affiliation{Department of Physics and National Center for Theoretical Sciences (NCTS), National Tsing Hua University, Hsinchu City, Taiwan 300.}
\affiliation{Donostia International Physics Center (DIPC), Manuel de Lardizabal, 4. E-20018 San Sebastian, Spain.}

\date{\today}

\begin{abstract}
We study the effect of anisotropy of the Rashba coupling on the extrinsic spin Hall effect due to spin-orbit active adatoms on graphene. In addition to the intrinsic spin-orbit coupling, a generalized anisotropic Rashba coupling arising from the breakdown of both mirror and hexagonal symmetries of pristine graphene is considered. We find that Rashba anisotropy can strongly modify the dependence of the spin Hall angle on carrier concentration. Our model provides a simple and general description of the skew scattering mechanism due to the spin-orbit coupling that is induced by proximity to large adatom clusters.
\end{abstract}

%\keywords{}

\maketitle

\section{\label{sec:level1}Introduction}

The spin Hall effect (SHE)\cite{Dyakonov_Perel_1,Dyakonov_Perel_2,hirsch1999spin,Zhang,Nagaosa_RevModPhys.82.1539} has been intensively investigated  in the last few decades due to its potential applications in spintronics.\cite{vzutic2004spintronics,Nagaosa_RevModPhys.82.1539} Generally speaking, the microscopic mechanisms of SHE can be classified into either  intrinsic~\cite{sinova2004universal,tanaka2008intrinsic} or extrinsic.\cite{tse2006spin,ferreira2014extrinsic} In both cases, the existence of spin-orbit coupling (SOC) in the material or heterostructure  is required. The intrinsic mechanism is a consequence of the band structure of the material whereas the extrinsic mechanism stems from scattering of the charge carriers by 
impurities that locally induce SOC. 

Since SOC is a relativistic effect that is typically strongest in materials containing heavy atoms,
the SOC in graphene~\cite{novoselov2005two,neto2009electronic} is expected to be 
weak.~\cite{huertas2006spin,min2006intrinsic,gmitra2009band} Therefore, graphene has been suggested as an ideal material for passive spintronics, for which a long spin diffusion length is required and SOC in the material is a major limiting factor.~\cite{you2015electron}

However, motivated by the search of materials exhibiting the quantum spin Hall effect,~\cite{KaneMeleQSHE1995} it has been theoretically predicted~\cite{weeks2011engineering,hu2012giant},  and experimentally  observed\cite{Balakrishnan:2014aa} that SOC can be greatly enhanced in graphene by means of adatom deposition. In the limit of a dilute number of impurities (i.e. adatoms), in which the excellent charge carrier mobility properties of graphene are not strongly modified, heavy adtom clusters have been predicted to induce a sizable SHE.~\cite{ferreira2014extrinsic}
Experimentally, a large spin Hall angle ($\sim 0.2$) has been reported by Balakrishnan and coworkers~\cite{Balakrishnan:2014aa} in devices made from chemical-vapor-deposited (CVD) graphene.  The phenomenon was explained~\cite{Balakrishnan:2014aa} by the combination of resonant scattering and the skew scattering off by residual Cu clusters resulting from the CVD process. It was experimentally estimated that the latter can induce a SOC of the order of $\sim 10$ meV.

Nevertheless, the nature of the SOC induced by the adatoms depends on their arrangement relative to hexagonal unit cell of graphene. The latter can lower the symmetry from the hexagonal symmetry of the carbon monolayer. It also depends on  the symmetry of the orbitals that hybridize with the $\pi$ bands of graphene, since this hybridization is ultimately responsible for both the proximity induced SOC and the resonant scattering. This also applies to heavy metal substrates or large clusters, where certain crystalline ordering is possible. In this regard, it is worth mentioning, for instance, the differences between gold intercalation ~\cite{marchenko2012giant}, which leads to a conventional Rashba splitting of graphene bands, and lead,~\cite{Calleja:2015aa} which results in a proximity induced Rashba SOC where the two terms of the coupling have different weight. The latter  is a consequence of the reduced orthorhombic symmetry of the composite (graphene + substrate) system. Such coupling is therefore an anisotropic generalization of the Rashba SOC, which arises due to the breakdown of both mirror and 6-fold rotation symmetry. Similar features have been reported recently in graphene intercalated with platinum.~\cite{Klimovskikh_etal_2015}

In this work, we shall investigate the skew scattering mechanism arising from the SOC induced by extrinsic scatterers. Unlike previous studies,~\cite{ferreira2014extrinsic,Balakrishnan:2014aa} we shall focus on the understanding of the effects of the Rashba anisotropy on the charge and especially spin transport 
properties, and in particular, the spin Hall angle.  To this end, we shall first solve the scattering problem of an anisotropic
SOC-active 
scatterer.~\cite{stauber2007electronic,ferreira2011unified}  From the single-impurity scattering data, we shall derive the relaxation times that parametrize the collision integral of the linearized Boltzmann transport equation (BTE)\cite{ziman1972principles}, which allows us to compute the spin Hall angle. Finally, we shall also compare the interplay and interference between different scattering potentials.

The rest of the article is organized as follows. We present the details of our theoretical model  in Sec.~\ref{sec:model}. First, we discuss the way the symmetries of  monolayer graphene decorated with adatoms constrain the form of the SOC in the $\mathbf{k}\cdot\mathbf{p}$ Hamiltonian of the system. Then, the single scatterer problem is solved. Using the scattering data (i.e. T-matrix) of the single scatterer problem, the linearized Boltzmann transport equation is also solved and the transport properties of the system are obtained.
In Sec.~\ref{sec:level3}, we discuss the most salient features of our results, namely the change of spin Hall angle and the conductivity of the system as a function of the anisotropy parameter.
A summary of the main results of this work is provided in Sec.~\ref{sec:concl}. Finally, let us mention that the Appendices contain the most technical details of the work.

\section{Model}\label{sec:model}

\subsection{Scattering potentials} \label{sec:level2}

 In  this study, we shall consider a \emph{dilute} ensemble of scatterers that create a (disorder) potential that is smooth in the atomic scale of graphene.  As a consequence, we shall neglect scattering between the two valleys at the opposite corners of the hexagonal Brillouin zone (i.e. $\mathbf{k} = \mathbf{K}_{\pm}$). Therefore, most of the discussion below applies to a single valley (i.e. $\mathbf{K}_{+}$) unless otherwise stated.
 
In order to understand the charge and spin transport properties of the system, we shall rely upon the semiclassical Boltzmann transport equation (BTE). The latter applies to doped graphene (i.e. when the Fermi energy measured from the Dirac point $E_F \neq 0$) in the limit where the distance between scatterers is much larger than the Fermi wavelength. Therefore, the results obtained from the BTE should be regarded as providing some sort of interpolation between the hole doped Fermi liquid ($E_F\ll 0$) and the electron doped Fermi liquid (i.e. $E_F \gg 0$ ) regime. 

In the \emph{dilute} impurity limit, the collision term of the BTE is determined by the scattering data for a single scatterer.~\cite{KohnLuttingerBTE} 
Thus, we first analyze the scattering problem
of a single scatterer, for which the $\mathbf{k}\cdot\mathbf{p}$ Hamiltonian describing the electron dynamics in the long-wavelength limit can be generally written as follows:
\begin{equation} \label{eq:H+V}
\mathcal{H}=\hbar v_F\left(\pm\sigma_x p_x+\sigma_y p_y\right)+ \sum_{\alpha=0,I,R}V_{\alpha}\left(\mathbf{r}\right),
\end{equation}
where the sign $\pm$ applies to the valley at crystal momentum $\mathbf{K}_{\pm}$ and $\sigma_{\alpha}$ ($\alpha = x,y,z$) are the Pauli matrices associated with the sublattice degrees of freedom of the wave function. The Pauli matrices acting on electron spin are denoted by $s_{\alpha}$. The first two terms in Eq. \eqref{eq:H+V} correspond to the  $\mathbf{k}\cdot \mathbf{p}$ Hamiltonian of the pristine graphene which describes the the electronic bands near the  $\mathbf{K}_{\pm}$ points.

Among the possible time-reversal invariant impurity potentials,  we shall focus on the scalar potential ($\alpha=0$), the intrinsic SOC ($\alpha=I$) and the Rashba  potential ($\alpha=R$). These three are invariant under the point group $C_{6v}$, which is generated by the 6-fold rotation axis perpendicular graphene intersecting the center of the hexagonal unit cell and 6 reflection planes containing such an axis. The latter group describes the rotation and mirror symmetries of monolayer graphene excluding the mirror reflection about the graphene plane which takes $z\to -z$. However, the scalar potential ($r = |\mathbf{r}|$), 
\begin{equation}\label{v0}
V_{0}(\mathbf{r})=\texttt{v}_0\left(r\right)\mathcal{I}_{4\times4},
\end{equation}
and intrinsic or Kane-Mele SOC term:
\cite{KaneMeleQSHE1995}
\begin{equation}\label{intrinsic}
V_{I}\left(\mathbf{r}\right)=\pm\Delta_{I}\left(\mathbf{r}\right)\sigma_z s_z.
\end{equation}
are invariant under the larger point group $D_{6h}$, which includes the mirror reflection for which $z \to -z$.  On the other hand, Rashba SOC is associated with the lack of the mirror reflection symmetry. Typically, this symmetry is broken and lowered to $C_{6v}$ by the presence of a substrate, adatoms, and/or ripples.

 The Rashba SOC is invariant under the $C_{6v}$ group
of pristine graphene and takes the following form:
\begin{equation}
V_{R}(\mathbf{r}) = \Delta_{R}(r) \left( 
 \pm\sigma_x  s_y-\sigma_y s_x \right). 
\end{equation}
However, in general,  the planar symmetry $C_{6v}$ can be broken, for example, due to the different symmetries of graphene lattice and the substrate, or the arrangement (relative to the hexagonal unit cell of graphene) of the adatoms in a large cluster in the proximity of the carbon layer. As a result, the symmetry can be lowered from hexagonal (i.e. $C_{6v}$) to rectangular (i.e. $C_{2v}$), for instance. This can be achieved by deposition or intercalation of a metal with either cubic or orthorhombic symmetry.\cite{Calleja:2015aa}  In 
Ref.~\onlinecite{Calleja:2015aa}, it was shown from symmetry arguments and first principle calculations that the two terms of the Rashba SOC can acquire different weights, which leads to an anisotropic form of the Rashba SOC potential:
\begin{equation}\label{Rashba}
V_{R}\left(\mathbf{r}\right)=\pm\Delta_1\left(r\right)\sigma_{x} s_{y}-\Delta_2\left(r\right)\sigma_{y} s_{x}.
\end{equation}
In this work, we shall study the effect of this anisotropy on the skew scattering mechanism and its contribution to the spin Hall effect.

 Before turning our attention to the study of the scattering problem by such anisotropic Rashba potential, it is useful to analyze the symmetries of~\eqref{eq:H+V} in the presence of the anisotropic Rashba SOC potential, Eq.~\eqref{Rashba}. For reasons that shall become clear below, it is convenient to write the anisotropic Rashba SOC as the sum of two terms, $V_{R}(r)=V_{SR} (r) + V_{NR}(r)$, where
\begin{equation}
V_{SR}( \mathbf{r} )=\Delta_{SR}(r) \left( \sigma_{x}  s_{y}- \sigma_{y}  s_{x} \right),
\end{equation}
\begin{equation}
V_{NR}(r)=\Delta_{NR}(r) \left( \sigma_{x}  s_{y} + \sigma_{y}  s_{x} \right),
\end{equation}
with
\begin{equation}
\Delta_{NR/SR}(r)=\frac{\Delta_{1}(r) \pm \Delta_{2}(r)}{2},
\end{equation}
where $+$ ($-$) sign applies to $\Delta_{NR}$ ($\Delta_{SR}$), respectively. For $\Delta_1 = \Delta_2$ we recover the standard Rashba SOC. In the opposite limit, $\Delta_1 = -\Delta_2$, a SOC to be termed `non-standard' Rashba is obtained.  This representation  enables us to  display more clearly how the  anisotropy in the Rashba 
SOC violates the conservation of the angular momentum projected onto the $z$ axis. Let us recall the  definition of the 
$z$-component of angular momentum operator:
\begin{equation}
J_{z}= l_{z} + \frac{\sigma_{z}}{2} +\frac{s_{z}}{2}.
\end{equation}
Notice that since $\Delta_{\alpha}(\mathbf{r})$ are functions  of $r = |\mathbf{r}|$, as we have assumed, the  scalar potential, intrinsic and standard Rashba SOC commute with $J_z$.  However,  when the Rashba SOC is anisotropic, $J_z$ is no longer conserved
and the culprit for this violation is the non-standard Rashba SOC introduced above.  Nevertheless, in the special case where $\Delta_1 = -\Delta_2$ (i.e. $\Delta_{SR} = 0$), the following quantity: 
\begin{equation}
M_{z}= l_{z} + \frac{\sigma_{z}}{2} - \frac{s_{z}}{2}, \label{eq:mz}
\end{equation}
is conserved instead of $J_z$. This will become useful in our investigation of this special limit for a more general type of scatterers than those considered in the following (see Appendix~\ref{app:nsr}).

%In what follows, we shall investigate spin transport in doped graphene decorated with adatom clusters, a situation that resembles the one studied in Refs.~\onlinecite{ferreira2014extrinsic,Balakrishnan:2014aa}.  However, unlike those studies, we shall take into account the possibility of anisotropy in the Rashba coupling.  

Note that the lack of conservation of $J_z$ by the anisotropic Rashba SOC makes it impossible to employ 
a partial wave expansion to solve the 
scattering problem as it was done in Ref.~\onlinecite{ferreira2014extrinsic}.
 Nonetheless, 
since we are interested in the scattering by clusters of adatoms of characteristic size $R \gg a$ ($a = 2.46\: \AA$ being the interatomic distance in graphene), and for typical experimental parameters in \emph{doped} graphene $R \ll k^{-1}_F$ (where
$k_F$ is the Fermi wave vector), we shall
approximate the cluster potentials 
by Dirac delta functions, i.e.,
\begin{align}
\texttt{v}_0\left(\mathbf{r}\right)&=\lambda_0\delta^{(2)}\left(\mathbf{r}\right),\label{eq:scpot}\\
\Delta_I\left(\mathbf{r}\right)&=\lambda_I\delta^{(2)}\left(\mathbf{r}\right),\\
\Delta_1\left(\mathbf{r}\right)&=\lambda_1\delta^{(2)}\left(\mathbf{r}\right),\\
\Delta_2\left(\mathbf{r}\right)&=\lambda_2\delta^{(2)}\left(\mathbf{r}\right).
\end{align}
Let us also define  $\text{v}_0 = \lambda_0 /R^2$, $\Delta_I = \lambda_I /R^2$ and $\Delta_{1,2} = \lambda_{1,2}/R^2$ as
the strength of the potentials in units of energy. In passing, we also note that a similar model (with only intrinsic SOC)  was successfully employed to account for the giant SHE observed in CVD graphene and attributed
to the SOC induced by residual Cu 
atom clusters.~\citep{Balakrishnan:2014aa}
Thus, the potentials in Eq.~\eqref{eq:H+V} take
the form:
\begin{equation}\label{eq:tot_soc}
V_{\alpha	}(\mathbf{r})= \lambda_{\alpha}\Lambda_{\alpha}\delta^{(2)}(\mathbf{r})
\; ; \; \alpha=\{0,I,SR,NR\},
\end{equation}
where $\lambda_{\alpha}$ is the strength of the potential and $\Lambda_{\alpha}$ are $4\times 4$ matrices acting upon the sublattice-spin degrees of freedom. Explicitly, the $\Lambda$ matrices are,
\begin{align}
\Lambda_{0}&=\mathcal{I}_{4\times 4}, \quad   \quad \Lambda_{1}=\sigma_{z}s_{z},  \\
\Lambda_{SR}&=\sigma_{x} s_{y}-\sigma_{y}s_{x} = i (\sigma^{-}s^{+} - \sigma^{+}s^{-}),   \\
\Lambda_{NR}&=\sigma_{x}s_{y}+\sigma_{y}s_{x} =  i (\sigma^{-}s^{-} - \sigma^{+}s^{+}).
\end{align}
When written in terms of $\sigma^{\pm}$ and
$s^{\pm}$, the conservation of $J_z$ by $V_{SR}$ and the failure to do so by $V_{NR}$ becomes apparent.

Interestingly, the $\Lambda$ matrices form a closed group under (matrix) multiplication. This means that the product of two of these matrices  can be written as a linear combination of 
\begin{equation}\label{eq:closure_relation}
\Lambda_i \Lambda_j = \sum_{l} c_{ijl} \Lambda_l.
\end{equation}
The coefficients  $c_{ijl}$ can be read off from table~\ref{tab:table1}. As a mathematical curiosity, it is worth noting that the group is abelian, as can be expected for a group of order four. Out of the two possible order four groups, this corresponds to the Klein group.  

\subsection{Single scatterer problem}\label{sec:level2-1} 

In this subsection, the Lippmann-Schwinger (LS) equation for the single impurity problem with the choice of potentials discussed in previous section will be solved. The LS  wave equation reads:
\begin{table}[b]
\caption{\label{tab:table1}%
The $\Lambda$ matrices span an order four group under matrix multiplication. Note that $\Lambda_{SR}$ is orthogonal to $\Lambda_{NR}$, i.e. the probability amplitude of being scattered consecutively by both standard Rashba SOC and non-standard Rashba SOC is zero.
}
\begin{ruledtabular}
\begin{tabular}{l  | c c c c }
  &
$\Lambda_{0}$&
$\Lambda_{I}$&
$\Lambda_{SR}$&
$\Lambda_{NR}$\\
\colrule
$\Lambda_{0}$ & $\Lambda_{0}$ &  $\Lambda_{I}$  & $\Lambda_{SR}$  &  $\Lambda_{NR}$ \\
$\Lambda_{I}$ &$\Lambda_{	I}$ & $\Lambda_{0}$ & $-\Lambda_{SR}$ & $\Lambda_{NR}$ \\
$\Lambda_{SR}$&$\Lambda_{SR}$& $-\Lambda_{SR}$  & $2(\Lambda_0-\Lambda_{I})$ & 0 \\
$\Lambda_{NR}$&$\Lambda_{NR}$& $\Lambda_{NR}$ & 0 & $2(\Lambda_0 + \Lambda_{I})$ 
\end{tabular}
\end{ruledtabular}
\end{table}
\begin{equation}\label{eq:1}
\Braket{{\bf r} | \psi_{\mathbf{p}}}=\Braket{{\mathbf{r}} | \phi_{\mathbf{k},\sigma}} +  \int d^{2}{\bf r'}\ \mathcal{G}_R({\bf r}-{\bf r')} 
\Braket{{\bf r'} | \mathcal{V}\left(\mathbf{r}\right) | \psi_{\mathbf{p}}},
\end{equation}
where
\begin{equation}
\mathcal{V}(\mathbf{r})=\sum_{\alpha=0,I,SR,NR}V_{\alpha}(\mathbf{r}),
\end{equation}
\begin{equation}
\Braket{{\bf r} | \phi_{\mathbf{k},\sigma}}=\frac{1}{\sqrt{2}}\left(\begin{array}{c}
1  \\
e^{i\theta_{k}} 
\end{array}\right) e^{i\bf k \cdot r} \eta_{\sigma},
\end{equation}
and $V_{\alpha}(\bf r)$ given by Eq.~\eqref{eq:tot_soc}. In the LS equation $\Braket{{\bf r} | \phi_{\mathbf{k},\sigma}}$ is the incident wave function from conduction band with momentum $\mathbf{p}$ and spin state described by the spinor $\eta_{\sigma}$, where 
$\eta_{\uparrow}=(1,0)^{T},  \eta_{\downarrow}=(0,1)^{T}$.
The spin quantization axis is taken to be the $z$ axis, which is perpendicular to the graphene plane. The angle of incidence is $\theta_k =\tan^{-1}\left(\frac{k_{y}}{k_{x}}\right)$; the normalization area of the system is taken to be unity. The scattered wave function is $\Braket{\mathbf{r}|\psi_{\mathbf{p} }}$. Note  that $\Braket{\mathbf{r}|\psi_{\mathbf{p} }}$ does not carry a spin index because it is not an eigenstate of $s_{z}$ in general. The (retarded) Green's function $\mathcal{G}_R({\bf r}-{\bf r'})$ is a $4\times 4$ matrix acting both in the sublattice pseudo-spin and electron-spin space (see Appendix~\ref{appA} for details). 

The Dirac-delta function potential allows us to express the solution to the LS equation in terms of the $\Lambda$ matrices,
\begin{align}
\Braket{{\bf r} | \psi_{\bf p}} =& 
\Braket{ {\bf r} | \phi_{\mathbf{k}, \sigma} } + 
\mathcal{G}_R({\bf r}) \sum_{i} \lambda_{i} \Lambda_{i} \Braket{{\bf 0}| \psi_{ \mathbf{p}}}\notag \\
 =& 
\Braket{ {\bf r} | \phi_{\mathbf{k}, \sigma} } + 
\mathcal{G}_R({\bf r}) \sum_{i,j} \lambda_{i} \beta_{j} \Lambda_{i} \Lambda_{j}\Braket{{\bf 0}| \phi_{ \mathbf{k},\sigma}},
\end{align}
where $\{i,j\}=\{I,0,SR,NR\}$. The coefficients $\beta_{i}$ are functions of the couplings $\lambda_{\alpha}$ that appear when we solve for 
$\Braket{{\bf 0} | \psi_{\bf p} }$ (see Appendix \ref{appA} for details).  In particular, it can be seen that for $\lambda_{NR}\rightarrow 0$, $\beta_{NR} \rightarrow 0$,  and it then follows that the scattered wave function is an eigenstate of $J_{z}$ because the expression for the scattered wave no longer contains $\Lambda_{NR}$. On the other hand, when $\lambda_{SR}\rightarrow 0$ then $\beta_{SR}\rightarrow 0$, the scattered wave function becomes an eigenstate of $M_{z}$ (cf. Eq.~\ref{eq:mz}). 
 
Using Eq.~\eqref{eq:closure_relation} and introducing the coefficients
\begin{equation}
\gamma_{l}=\sum_{i,j}c_{ijl}\lambda_{i}\beta_{j},
\end{equation}
the solution to the LS equation \eqref{eq:1}  takes the following 
compact form:
\begin{equation}
\Braket{{\bf r} | \psi_{\bf p}} =
\Braket{ {\bf r} | \phi_{\mathbf{k}, \sigma} } + 
\mathcal{G}_R({\bf r}) \sum_{l} \gamma_{l} \Lambda_{l} 
\Braket{{\bf 0}| \phi_{ \mathbf{k},\sigma}}.
\end{equation}
Note that the right hand-side of the above expression contains only known quantities. After expanding the Green function asymptotically at large distances, the scattered wave can be written as the sum of an incident and an outgoing wave:
\begin{equation}
\Braket{{\bf r} | \psi_{\mathbf{p}}} \approx
\Braket{{\bf r} | \phi_{\mathbf{k},\sigma }} + \sum_{\sigma'=\uparrow,\downarrow} 
f(\mathbf{p} , \sigma' ; \mathbf{k},\sigma ) \frac{e^{ikr} }{\sqrt{2r} }  \left(\begin{array}{c}
1  \\
e^{i\theta_{k}} 
\end{array}\right) \eta_{\sigma'}.
 \end{equation}
In the above expression we have introduced the scattering amplitudes given by $f(\mathbf{k},\sigma; \mathbf{p} , \uparrow)$ and $f(\mathbf{k},\sigma ; \mathbf{p} , \downarrow)$. From it, the differential scattering cross-section can be calculated using:
  \begin{align} \label{eq:differential_cross_section}
  \frac{d\sigma}{d\theta} =&\sum_{\sigma'=\uparrow,\downarrow} 
\Big | f(\mathbf{p} , \sigma' ; \mathbf{k},\sigma ) \Big|^{2} \\
=& \sum_{\sigma'=\uparrow,\downarrow} 
\Big | f_{\sigma' \sigma}(\theta) \Big|^{2},
  \end{align}
where $\theta =\cos^{-1}\left(\frac{\bf k \cdot p}{ k^2}\right)$ is the scattering angle. We refer the reader to the Appendix \ref{appA}  for the detailed form of the scattering amplitude and how it is related with the scattering T-matrix that enters in the collision term of the Boltzmann transport equation. 

\subsection{Transport properties}

In order to compute the charge and spin transport properties of a dilute \emph{random} ensemble of identical clusters of areal density $n_{imp}$, we use the semi-classical Boltzmann transport equation (BTE).~\cite{ziman1972principles,ferreira2014extrinsic}  The details of its solution in the linearized approximation are reviewed in Appendix~\ref{appAB}. The exact solution of this equation~\citep{ferreira2014extrinsic}  allows us to obtain the charge,  and spin Hall conductivities ($\sigma_{tr}$ and $\sigma_{sH}$, respectively):
\begin{eqnarray}
\sigma_{tr}=\frac{e^2}{h}\int d\epsilon \frac{| \epsilon |}{\hbar} \left( \frac{\partial n }{\partial{\epsilon}} \right) \frac{\tau_{sk}\tau_{sk}^* \tau_{tr}}{\tau_{sk}\tau_{sk}^* +\tau_{tr}\tau_{tr}^*},\label{eq:6} \\
\sigma_{sH}=-\frac{e^2}{h}\int d\epsilon \frac{| \epsilon |}{\hbar} \left( \frac{\partial n}{\partial{\epsilon}}\right) \frac{\tau_{sk}\tau_{tr}^* \tau_{tr}}{\tau_{sk}\tau_{sk}^* +\tau_{tr}\tau_{tr}^*}, \label{eq:7}
\end{eqnarray}
where $n(\epsilon) = (e^{(\epsilon-\mu)/k_B T}+1)^{-1}$ is the Fermi-Dirac distribution. The different scattering times, $\tau_{tr},\tau_{sk},\tau_{tr}^*,\tau^*_{sk}$ are defined in Appendix~\ref{appAB} and can be derived from the differential scattering cross section. In particular, we would like to point out that
\begin{align}
\frac{1}{\tau^*_{tr}}&= n_{imp}v_F\sum_{\sigma'=\uparrow, \downarrow} \int  (1-\eta_{\sigma\sigma'}\text{cos} \theta) 
\Big | f_{\sigma' \sigma}(\theta) \Big|^{2} d\theta \nonumber\\
&\equiv n_{imp} v_F \Sigma^*_{tr},
\end{align}
where $n_{imp}$ and $v_{F}$ are the areal density of the impurities and the Fermi velocity of pristine graphene, respectively. Here, the notation $\eta_{\uparrow,\downarrow}=\eta_{\downarrow,\uparrow}=-1$ and $\eta_{\uparrow,\uparrow}=\eta_{\downarrow,\downarrow}=+1$. $\Sigma_{tr}^{*}$ is the single-scatterer transport cross-section which exhibit sharp peaks at the resonance energies of the single scatterer.

The figure of merit determining the efficiency in the charge current to spin current conversion, known as spin Hall ``angle", is defined as the ratio:
\begin{equation}
\gamma=\frac{\sigma_{sH}}{\sigma_{tr}}.
\end{equation}
At zero temperature, the spin Hall angle reduces to the ratio (see Appendix \ref{appAB}):
\begin{equation}
\gamma=-\frac{\tau_{tr}^*}{\tau_{sk}^*}.
\end{equation}

%Boltzmann, spin Hall angle%

\begin{figure}[hbtp]
\centering
 \includegraphics[scale=.55]{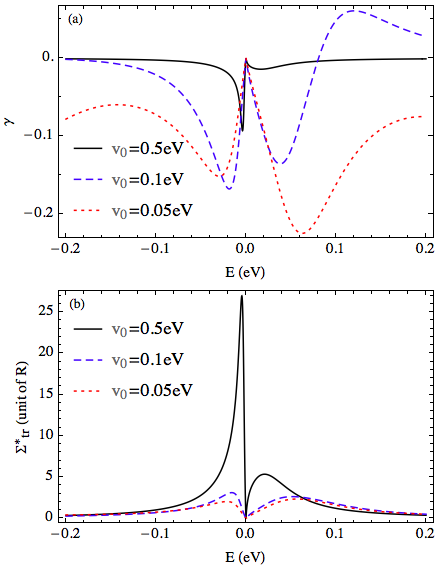}
\caption{(a) Spin Hall angle $\gamma$ versus energy of the carrier ($E$) for different values of the strength of scalar potential $\text{v}_0 = \lambda_0 /R^2$.  We set $R=20$ nm, $T=0$ K, $\Delta_I= \lambda_I /R^2 = 15$ meV, $\beta=-\frac{\pi}{8}$ and  $\frac{\Delta_1+\Delta_2}{2} = \frac{(\lambda_1+\lambda_2)}{2R^2}=15$ meV. The location of resonances gradually approaches the Dirac point when v$_0$ increases, illustrating the general feature of resonant scattering even for the anisotropic case. (b) Corresponding starred transport cross sections of each case. See App. \ref{appAB} for the definition of starred transport cross section. }
\label{changev0}
\end{figure}

\begin{figure}[hbtp]
\centering
\includegraphics[scale=.55]{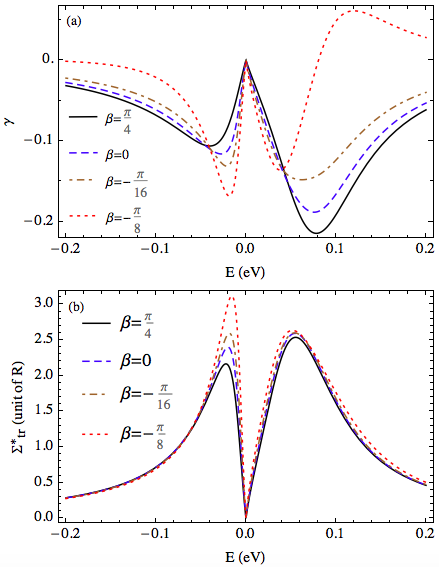}
\caption{(a) Spin Hall Angle $\gamma$ versus the energy of the scattered electron $E$ for different anisotropy values of the Rasha SOC anisotropy parameter $\beta = \tan^{-1}\left(\lambda_2/\lambda_1\right)$. We have chosen the strength of the scalar part of the potential, $\text{v}_0 = \lambda_0/R^2 =  0.1$ eV and other parameters are the same as Fig. \ref{changev0}. The magnitude of the spin Hall angle  $\gamma$ is enhanced near the position of the scattering resonance, which is signalled by a peak in the transport  scattering   cross section (b) Corresponding starred transport cross sections $\Sigma^*_{tr}$ of each case. The curves for the transport 
cross section and the spin Hall angle are asymmetric about 
$E = 0$ because the Rashba SOC breaks the particle-hole symmetry of the $\mathbf{k}\cdot\mathbf{p}$ Hamiltonian.}
\label{changebeta}
\end{figure}

\begin{figure}[hbtp]
\centering
\includegraphics[scale=.42]{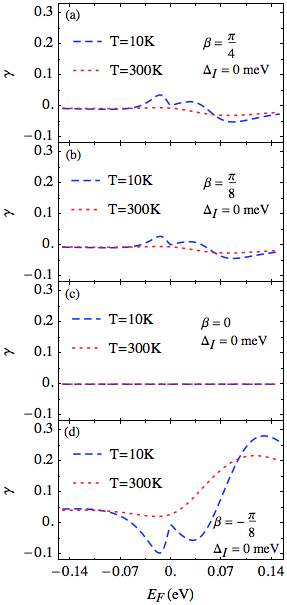}
\includegraphics[scale=.42]{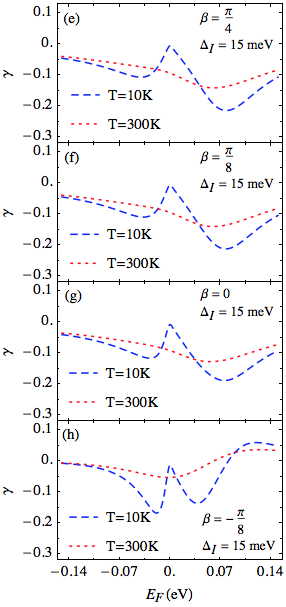}
\caption{(a)-(d) Temperature dependence of $\gamma$ versus $E_F$ plot at $\Delta_I=0$meV and four different $\beta$. (e)-(h) Temperature dependence of $\gamma$ versus $E_F$ plot at $\Delta_I=15$meV and four different $\beta$. v$_0=0.1$eV in all the eight plots. Other parameters are the same as in Fig.\ref{changebeta}. and Fig. \ref{changev0}.}
\label{temp}
\end{figure}

\section{Results and discussion}\label{sec:level3}

 The following discussion is about the effects of the anisotropy in the Rashba SOC. The degree of anisotropy in  Rashba SOC will be phrased in terms of the anisotropy parameter:
\begin{equation}
\beta = \tan^{-1}\left(\frac{\Delta_2}{\Delta_1}\right) = \tan^{-1} \left(\frac{\lambda_2}{\lambda_1}\right).
\end{equation}
Note that for the standard (``isotropic'') Rashba SOC, 
$\beta = \frac{\pi}{4}$. For $\beta = -\frac{\pi}{4}$,
the Rashba-like SOC is of the form 
$\sim \left(\tau \sigma_x s_y + \sigma_y s_x \right)$ (with $\tau=\pm$ for $\mathbf{K}_{\pm}$), which 
has been termed non-standard Rashba in Sec,~\ref{sec:level2}.  
In the range $-\frac{\pi}{4}<\beta<\frac{\pi}{4}$,   $\beta$ 
is a measure of the degree to which  the  $C_{6v}$ symmetry is broken
by  the adatom arrangement within the clusters. Close to $\beta  = \frac{\pi}{4}$ the deviation from Rashba and the perfect $C_{6v}$ symmetric situation is small. On the other hand, having $-\frac{\pi}{4}<\beta < 0$ requires a strong breaking of the  $C_{6v}$ symmetry.

In Fig.~\ref{changev0}, we show the dependence of the spin Hall angle $\gamma$  and the transport cross section $\Sigma^*_{tr}$ at zero temperature on the carrier energy $E$  for different values of strength of the scalar potential $\text{v}_0 =\lambda_0/R^2$ (cf. Eq.~\ref{eq:scpot}) for a rather anisotropic Rashba-like SOC corresponding to $\beta = -\pi/8$. It can be seen that the enhancement of the spin Hall angle still takes place around the values of $E$ for which $\Sigma^*_{tr}(E)$ exhibits a peak, that is, a scattering resonance. This is in agreement with what was already pointed out in Ref.~\onlinecite{ferreira2014extrinsic} for the isotropic Rashba SOC. Physically, this is also expected, because at resonance the 
scattering electron or hole spends most time near the 
scatterer and therefore it can also experience the effect 
of the locally induced SOC.  The enhancement of $\gamma$ is suppressed at large values of $\text{v}_0$. To understand this effect qualitatively, let us recall that $\gamma$ and $\Sigma^*_{tr}$ are both determined by the T-matrix, which obeys the LS equation:
\begin{equation}
T(E) = V + V\mathcal{G}_R(E)T(E), \label{eq:tls}
\end{equation}
where $\mathcal{G}_R(E)$ is the retarded Green's function and $V = V_0 + V_{SOC}$, $V_0\propto \text{v}_0$ being the scalar potential and $V_{SOC} = V_{I} + V_R$ the  SOC part of  the potential. In the limit where $V_0 \gg V_{SOC}$, the solution to Eq.~\eqref{eq:tls} can be (loosely) written as:
\begin{equation}
T(E) = \frac{V_0 + V_{SOC}}{1 - (V_0+V_{SOC})\mathcal{G}_R(E)}\approx \frac{-1}{\mathcal{G}_R(E)} \left[1 + \frac{V_{SOC}}{V_0} \right],\label{eq:approx}
\end{equation}
where the last expression applies to the large $V_0$ limit. 
Thus,  to leading order, the
cross section $\Sigma^*_{tr}$ is determined by the
first term of the right hand-side, whereas
$\gamma$ is determined by the the second term. Hence, $\gamma$
is expected  to decrease at large $V_0 \propto \text{v}_0$,
as shown in Fig.~\ref{changev0}.

For a given set of $\text{v}_0$, $\Delta_I$, and $\Delta_{SR}$, Fig.~\ref{changebeta} shows the behaviour of $\gamma$ and $\Sigma^*_{tr}$  as a function of the incident electron energy $E$ at different values of anisotropy parameter $\beta$. For the values of $\beta$ close to those corresponding to the non-standard Rashba SOC (i.e. for $\beta \approx -\frac{\pi}{4}$), the energy dependence is strongly modified. On the other hand,  for the case of a delta function potential, the anisotropy has a less pronounced effect on the cross section $\Sigma^*_{tr}$.

The observations made above remain largely unchanged  when the effect of finite temperature is taken into account, see Fig~\ref{temp}. As shown there, thermal fluctuations and the associated smearing of the Fermi distribution,  smooth out the sharper features of the (Fermi) energy dependence $E_F$ of $\gamma$ found at $T = 0$ and suppress the magnitude of $\gamma$. This can be seen 
in the left panel in Fig.~\ref{temp} for case of a pure (i.e. $\Delta_I =0$) anisotropic Rashba SOC and on the right for $\Delta_I \neq 0$. The plots on the left panel also illustrate that, $\beta = 0$ (i.e. $\Delta_2 = 0$) the spin current as well as the spin Hall angle vanish (cf.  second plot from the bottom on the left).  This is because the quantization axis for the spin current is aligned along the $z$ axis,  whereas for $\beta = 0$, $s_y$ commutes with
the Hamiltonian.  As pointed out above for $T =0$, the energy dependence (relative to the isotropic case), is most strongly affected as $\beta$ approaches $-\frac{\pi}{4}$ (see plot for $\beta = -\frac{\pi}{8}$). However,  the effect of the anisotropy is less pronounced for $\beta > 0$. This conclusion still holds true when the scatterer also induces intrinsic SOC on the graphene layer (i.e. for $\Delta_I \neq 0$), as it is shown in the right panel of Fig.~\ref{temp}. 

\begin{figure}
\centering
\includegraphics[scale=.53]{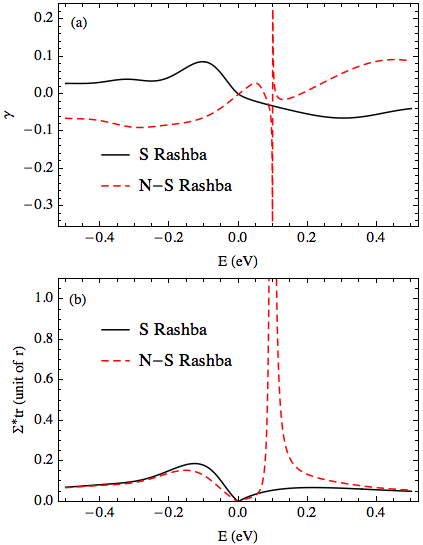}
\caption{(a) Comparison between the spin Hall angle ($\gamma$) of a standard Rashba (SR) and non-standard Rashba (NS) SOC for a circular (i.e. `pill-box') scattering potential. (b)  Transport cross section. We have assumed the radius of the scatterer $R=4$ nm, $\text{v}_0=0.1$eV and $\Delta_{SR}=\Delta_{NR}=25$meV. The same asymmetric resonant property as that discussed in the main context recurs in the finite-disk modeling.}
\label{SR and NR comparison}
\end{figure}

Finally, it is also worth mentioning that the observation of a very different
energy dependence as $\beta \to  -\frac{\pi}{4}$ is independent of the assumption of a Dirac delta potential. This is investigated in detail in Appendix~\ref{appC}, where circular (i.e. `pill-box' shaped) scatterer is assumed and the scattering properties in the the case of standard and non-standard Rashba are obtained.
The results for the energy dependence of $\gamma$ and $\Sigma^*_{tr}$ are displayed in Fig.~\ref{SR and NR comparison}. The more complicated 
internal  structure of the finite-radius circular scatter, 
whose wave functions are distorted in different ways by the standard and non-standard Rashba SOC, shows up in a very different resonant peak structure
exhibited by the transport cross section $\Sigma_{tr}^{*}$ and the spin Hall angle $\gamma$. 
\section{Summary and Conclusions}\label{sec:concl}

We have analyzed a simple model to 
understand the effects of the anisotropy of the proximity-induced Rashba spin-orbit coupling (SOC) on the spin Hall effect. The anisotropy arises as a consequence of the arrangement of adatoms in the clusters decorating a single layer of graphene and takes the form:
\begin{equation}
V_R = \Delta_1(r) \sigma_x s_y - \Delta_2(r) \sigma_y s_x.
\end{equation}
On symmetry grounds, such a SOC is effectively generated when the arrangement lowers the symmetry of the system from the hexagonal symmetry (i.e. the $C_{6v}$ group) of graphene.

From our analysis we conclude that  the anisotropy in the Rashba SOC does not modify the observation that the spin Hall angle in graphene is enhanced by the scattering resonances~\cite{ferreira2014extrinsic} that appear near the Dirac point. In addition, the dependence on the carrier concentration (or equivalently the Fermi energy) of the spin Hall angle is also not strongly modified for weak anisotropy. However, when the parameter  
 $\beta = \tan^{-1}(\Delta_2/\Delta_1) < 0$ and especially when $\beta$ approaches $-\frac{\pi}{4}$,  we have found the Fermi energy dependence to strongly deviate from the one observed in the isotropic case (corresponding to $\Delta_1 = \Delta_2$ or $\beta = \frac{\pi}{4}$). This conclusion is robust against finite temperature effects, which somewhat smoothes out the Fermi energy dependence and suppress the value of $\gamma$. It is also not modified by relaxing our assumption of a zero-range (i.e. Dirac-delta) potentials.

 In our study, we assumed a single type of single scatter model. In a realistic
experiment (like the one envisaged in Ref.~\onlinecite{Balakrishnan:2014aa}), several
kinds of scatterers may be present, some of which do not induce SOC. However,
we expect that the above qualitative features will remain unchanged. Experimentally, it would be interesting to study the differences in the Fermi energy (i.e. doping) dependence of the spin Hall angle for clusters of different atomic species, which can lead to different anisotropic Rashba couplings. Indeed, experimental evidence for intercalated Pb islands 
obtained in Ref.~\onlinecite{Calleja:2015aa} seems
to indicate that this metal can induce a rather anisotropic Rashba coupling with $\Delta_2\gg \Delta_1$. Similar deviations from the standard Rashba splitting of graphene bands have been recently reported in platinum intercalated devices.\cite{Klimovskikh_etal_2015} Our study identifies the signatures of such deviations in the carriers' skew scattering properties, providing a way to probe different spin textures in transport.

\acknowledgements

C.H's work was supported in part by the Singapore National Research
Foundation grant No.~NRFF2012-02.
HYY and MAC acknowledge support from the Ministry of Science and Technology in Taiwan. HO acknowledges support from the European Union's Seventh Framework Programme (FP7/2007-2013) through the ERC Advanced Grant NOVGRAPHENE (GA No. 290846). 

\appendix
\section{Single scatterer problem}\label{appA}

In this Appendix, we shall provide the details of the solution of the 
Lippmann-Schwinger (LS) equation. To this end, we first recall the form of the 
 retarded Green's function in the continuum limit of single-layer graphene, 
 $\mathcal{G}_{R}(\mathbf{r -r'}) = G_{R}(\bf r -r')\otimes \mathbf{I}$, where  $\mathbf{I}=\sum_{\sigma=\uparrow,\downarrow} \eta_{\sigma} \eta_{\sigma}^{\dagger}$  is the identity matrix in spin space.
%with 
%
%\begin{equation}
%\eta_{\uparrow}=(1,0)^{T}\; ; \;  \eta_{\downarrow}=(0,1)^{T},
%\end{equation}
%
The function $G_{R}(\mathbf{r -r'})$ is given by
\begin{equation}
G_{R}({\bf r},{\bf r'})  =\Braket{{\bf r} | \frac { 1 }{ E+ i0^{+}-H_0} | {\bf r'}}. 
% \nonumber\\
% =\Bra{{\bf r}}(E-{\bf \sigma \cdot \hat{p}}) \hat{1}_{{\bf k'}} (\frac{1}%
%{E^2-{\bf \sigma \cdot \hat{p}}^2}) \hat{1}_{{\bf k''}}\Ket{{\bf  r}}
\end{equation}
Hence,
%
%\begin{align}
%G_{R}({\bf r},{\bf r'}) & =\sum_{{\bf k}}(E-i\hbar v_F {\boldsymbol{\sigma} %
%\cdot \boldsymbol{\nabla}_{\bf r}})  \frac{\Braket{{\bf r} | {\bf k}}\Braket{{\bf k} | {\bf r'}}}{\left(E+ i0^{+}\right)^2-\hbar^2v_F^2 \left|\mathbf{k}\right|^2}.
%\end{align}
%
%Turning the sum into an integral, we have
%
\begin{align}\label{eq:A3}
G_{R}({\bf r},{\bf r'})&=
(E-i\hbar v_F \boldsymbol{\sigma} \cdot \boldsymbol{\nabla}_{\bf r})\cdot
\nonumber\\
& \int \frac{d^2{\bf k}}{(2\pi\hbar v_F)^2}\frac{e^{i{\bf k'} \cdot ({\bf r}-{\bf r'})}}{\left(\frac{E}{\hbar v_F}+ i0^+\right)^2-\left|\mathbf{k}\right|^2}.
\end{align}
The integral in the above expression is the Green's function for the  two-dimensional Helmholtz equation, which reads:
\begin{equation}
G^H_R({\bf r}-{\bf r'})=-\frac{i}{4 \hbar^2v_F^2} H^{(1)}_0\left( \frac{E|{\bf r}-{\bf r'}|}{\hbar v_F}\right),
\end{equation}
where $H^{(1)}$ is the Hankel function of the first kind. Inserting this result in the expression for $G_{R}({\bf r},{\bf r'})$ and using  $\frac{d}{dz}H^{(1)}_m(z)=\frac{mH^{(1)}_m(z)}{z}-H^{(1)}_{m+1}(z)$ yields:
\begin{align}
G_{R}({\bf r}-{\bf r'})&=\frac{-i|E|}{4\hbar^2 v_F^2}\left[\text{ sign}(E)H^{(1)}_0\left( \frac{E|{\bf r-r'}|}{\hbar v_F}\right)
\right.\nonumber\\
&\left.  +i\sigma_{\theta} H^{(1)}_1\left(\frac{E|{\bf r-r'}|}{\hbar v_F}\right)\right],
\end{align}
where 
\begin{equation}
\sigma_\theta=
\left(\begin{array}{cc}
0 & e^{-i\theta} \\
e^{i\theta} & 0 
\end{array}\right).
\end{equation}
Due to translational invariance, the Green function is only a function of the difference in position. Here $\theta$, with $\cos \theta = (\mathbf{r}  - \mathbf{r'}) \cdot \hat{x}$, is the angle between the vector ${\bf r-r'}$ and the x-axis.  Note that we have chosen the retarded Green's function for both electrons (i.e. $E>0$) and holes (i.e. $E<0$). Setting $|E|=\hbar v_F k$, we arrive at
\begin{widetext}
\begin{equation}
G_R\left(\mathbf{r} - \mathbf{r}'\right)=\begin{cases}
\frac{-ik}{4 \hbar v_F}\left[H^{(1)}_0\left(k|{\bf r-r'}|\right)+i\sigma_{\theta} H^{(1)}_1\left(k|{\bf r-r'}|\right)\right]& \text{for }E>0,  \\
\frac{ik}{4 \hbar v_F}\left[-H^{(2)}_0\left(k|{\bf r-r'}|\right)+i\sigma_{\theta} H^{(2)}_1\left(k|{\bf r-r'}|\right)\right]& \text{for }E<0.
\end{cases}
\end{equation}
\end{widetext}
To simplify the discussion, in what follows, we shall limit ourselves to the study of the scattering of electrons within the conduction band (i.e. for $E > 0$), although in the main text both the valence ($E<0$) and conduction ($E > 0$) bands have been considered. 

As described in the Sec.~\ref{sec:level2-1}, in order to obtain the asymptotic wave function describing the outgoing scattered wave, we need to consider the limit  
$|\mathbf{r}| \gg |\mathbf{r}^{\prime}|$, in which the Green's function becomes:
\begin{equation}
G_R\left(\mathbf{r} - \mathbf{r}'\right) \approx
\frac{-ik}{4 \hbar v_F} \sqrt{\frac{2}{\pi k r}} e^{i(kr-\frac{\pi}{4})}  e^{-i \bf p \cdot \bf r'}
\left(\begin{array}{cc}
1 & e^{-i\theta_{p}} \\
e^{i\theta_{p}} & 1 
\end{array}\right),
\end{equation}
\begin{equation}
\mathbf{p} \equiv k \frac{\bf r}{| \mathbf{r} |},
\end{equation}
where $\mathbf{p}$ is the momentum of the scattered wave and $\cos \theta_{p}=\cos \theta_{r}=\frac{\mathbf{r}}{|\mathbf{r}|}\cdot \hat{x}$ is the angle subtended between the scattered momentum with $x$ axis. 
%The momentum $\bf p$ has the same magnitude as the incident wave (elastic scattering) and points in the direction where the scattered wave function is observed. 

Accounting for the spin degree of freedom, the asymptotic form of  Green's function reads:,
\begin{align}
\mathcal{G}_R\left(\mathbf{r} - \mathbf{r}'\right) \approx & 
\frac{-ik}{4 \hbar v_F} \sqrt{\frac{2}{ \pi k r}} e^{i(kr-\frac{\pi}{4})} \sum_{\sigma}  \left(\begin{array}{c}
1  \\
e^{i\theta_{p}} 
\end{array}\right) \eta_{\sigma}  \notag \\ 
& \eta_{\sigma}^{\dagger} \left(1\; e^{-i\theta_{p}} \right) e^{-i \bf p \cdot \bf r'}.
\end{align}

Thus, we are equiped to solve the Lippmann-Schwinger  (LS) equation of the scattering problem.  Assuming an incident electron from the conduction band with momentum $\mathbf{p}$ and $\sigma$ means that (we work assuming a normalization area equal to unity):
\begin{equation}
\Braket{{\bf r} | \phi_{\mathbf{k},\sigma}}=\frac{1}{\sqrt{2}}\left(\begin{array}{c}
1  \\
e^{i\theta_{k}} 
\end{array}\right) e^{i\bf k \cdot r} \eta_{\sigma}.
\end{equation}
Thus, the LS equation ( Eq. \ref{eq:1} ) becomes
\begin{widetext}
\begin{align}
\Braket{{\bf r} | \psi_{\mathbf{p}}}= & 
\Braket{{\bf r} | \phi_{\mathbf{k},\sigma}} + \sum_{\sigma'=\uparrow,\downarrow}
\frac{-ik}{4 \hbar v_F} \sqrt{\frac{2}{ \pi k r}} e^{i(kr-\frac{\pi}{4})}  
\left(\begin{array}{c}
1  \\
e^{i\theta_{p}} 
\end{array}\right) \eta_{\sigma'}
\int d^{2}{\bf r'}  \left(1\; e^{-i\theta_{p}} \right) \eta_{\sigma'}^{\dagger}
 e^{-i \bf p \cdot \bf r'} 
\Braket{{\bf r'} | T | \phi_{\mathbf{k}, \sigma} } \nonumber \\
= &
\Braket{{\bf r} | \phi_{\mathbf{k},\sigma }} + \sum_{\sigma'=\uparrow,\downarrow} 
f(\mathbf{p} , \sigma' ; \mathbf{k},\sigma ) \frac{e^{ikr} }{\sqrt{2 r} }  
 \left(\begin{array}{c}
1  \\
e^{i\theta_{p}} 
\end{array}\right) \eta_{\sigma'} ,
 \end{align}
\end{widetext}
where we have introduced the T-matrix, which can be defined by the equation $T | \phi_{\mathbf{k},\sigma}\rangle = \mathcal{V} | \psi_\mathbf{p} \rangle$.  Note that the scattered wave does not carry a spin index because it is not an eigenstate of $s_{z}$. %Furthermore, it is not normalized yet. 
Indeed, for a non-spin conserving potential like Rashba,  the scattered wave is a combination of the incident wave with momentum $\bf k$ and spin $\sigma$ and an scattered radial spin up (spin down) wave
 with amplitude given by $f(\mathbf{k},\sigma; \mathbf{p} , \uparrow)$
  ($f(\mathbf{k},\sigma ; \mathbf{p} , \downarrow)$). %It should be clear from the above equation that the scattered wave is in the conduction band.
  
From the above result, the scattering amplitude can be related to the T-matrix by the following expression:
\begin{equation}\label{eq:Tmatrix_scat_amp}
f(\mathbf{p} , \sigma' ; \mathbf{k},\sigma )=   \frac{-  i e^{-i\frac{\pi}{4}} }{\hbar v_F}  \sqrt{\frac{k}{2\pi}}
\langle \phi_{\mathbf{p},\sigma'} |T|  \phi_{\mathbf{k},\sigma} \rangle.
\end{equation}
For elastic scattering (i.e. $|\mathbf{p}|=|\mathbf{k}|$), the scattering amplitude is a function of the scattering angle $\theta =\arccos\frac{\bf k \cdot p}{ k^2}$, i.e.
\begin{equation}\label{eq:elastic}
f(\mathbf{p} , \sigma' ; \mathbf{k},\sigma )=f_{\sigma',\sigma}(\theta).
\end{equation}
Recalling that the impurity potential is given by Eq.~\eqref{eq:tot_soc}, the T-matrix can be written as follows:
\begin{align}
\langle \phi_{\mathbf{p},\sigma'} |T|  \phi_{\mathbf{k},\sigma} \rangle =&
\int d^{2}r \Braket{ \phi_{\mathbf{p},\sigma'}| \bf r} \Braket{ \mathbf{r}| 
V(\mathbf{r})|  \psi_{\mathbf{p}}} \\
=& \sum_{i } \lambda_{i} \Braket{ \phi_{\mathbf{p},\sigma'}| \bf 0} \Lambda_{i}\Braket{ \mathbf{0}| \psi_{\mathbf{p}}} \\
=&\sum_{i,j} \lambda_{i}\beta_{j} \Braket{ \phi_{\mathbf{p},\sigma'}| \bf 0} \Lambda_{i} \Lambda_{j} \Braket{ \mathbf{0}| \phi_{\mathbf{k},\sigma}},
\end{align}
where we have used (see below) $
\Braket{{\bf 0} | \psi_{\mathbf{p}}}= \sum_{j}\beta_{j}\Lambda_{j}\Braket{{\bf 0} | \phi_{\mathbf{k},\sigma}}
$.
Writing $ \Lambda_i \Lambda_j = \sum_{l} c_{ijl} \Lambda_l $ where the coefficient $c_{ijl}$ can be read off from group multiplication table (cf.~\ref{tab:table1}). In addition, let us define:
\begin{equation}
\gamma_{l}=\sum_{i,j}c_{ijl}\lambda_{i}\beta_{j}.
\end{equation}
The T-matrix can be obtained as follow,
\begin{equation}
\langle \phi_{\mathbf{p},\sigma'} |T|  \phi_{\mathbf{k},\sigma} \rangle =\sum_{l} \gamma_{l} \Braket{ \phi_{\mathbf{p},\sigma'}| \bf 0} \Lambda_{l} \Braket{ \mathbf{0}| \phi_{\mathbf{k},\sigma}}.
\end{equation}
In the above equations, it is understood that the indices $i,j,l$ run over the set $\{0,I,SR,NR\}$. % \textcolor{red}{Upon setting ${\bf x=0}$ in Eq.~\eqref{eq:1}, we get that  $\Braket{{\bf 0} | \psi_{\mathbf{p}}}= \sum_{j}\beta_{j}\Lambda_{j}\Braket{{\bf 0} | \phi_{\mathbf{k},\sigma}} $  and therefore,}
Upon setting  ${\bf r =0}$ in Eq.~\eqref{eq:1}, 
\begin{align}
\Braket{{\bf 0} | \psi_{\mathbf{p}}} & =\Braket{{\bf 0} | \phi_{\mathbf{k},\sigma}}+\sum_{j} \mathcal{G}_R({0}) \Lambda_{j} \Braket{{\bf 0}| \psi_{\mathbf{p}}}.
\end{align}
Hence,
\begin{align}
\Braket{{\bf 0} | \psi_{\mathbf{p}}}&= \frac{1}{1- \sum_{j} \mathcal{G}_R(0) \Lambda_{j}}\Braket{{\bf 0} | \phi_{\mathbf{k},\sigma}} \\
&= \sum_{j}\beta_{j}\Lambda_{j}\Braket{{\bf 0} | \phi_{\mathbf{k},\sigma}}.
\end{align}
The coefficients $\beta_i$ are obtained by inverting the matrix $1 -  \sum_{j} \mathcal{G}_R(0) \Lambda_j$ and (exactly) projecting onto the basis of $\Lambda$ matrices, 
which yields: 
\begin{widetext}
\begin{align}
& \beta_0 =\frac{1}{4} \sum_{\eta=\pm }\sum_{\eta'=\pm }\left( \frac{1}{1+G_R(0)(\lambda_{I}+2\eta\lambda_{SR}-\eta' \lambda_0)} \right),\\
& \beta_I =\frac{1}{4} \sum_{\eta=\pm }\sum_{\eta'=\pm }\left( \frac{\eta'}{1+G_R(0)(\lambda_{I}+2\eta\lambda_{SR}-\eta' \lambda_0)} \right),\\
& \beta_{SR} = -\frac{G_R(0) \lambda_{SR}}{(1+G_R(0)(\lambda_I-2\lambda_{SR}- \lambda_0))(1+G_R(0)(\lambda_I+2\lambda_{SR}-\lambda_0))}, \\
& \beta_{NR} = -\frac{G_R(0) \lambda_{NR}}{(1-G_R(0)(\lambda_I-2\lambda_{NR}+\lambda_0))(1-G_R(0)(\lambda_I+2\lambda_{NR}+ \lambda_0))}.
\end{align}
\end{widetext}
In the above expressions $G_{R}(0)$ is a scalar and 
 $\mathcal{G}_R({0})=G_{R}(0)\otimes \mathbf{I}$ is the matrix Green's function at the origin. The $G_{R}(0)$ is obtained by the imposing a cut-off at high momenta.  Setting ${\bf r=r'=0}$ in Eq.~\eqref{eq:A3}, we have
\begin{align}
G_R(0)= \int \frac{d k'}{2\pi\hbar v_F}\frac{k\text{ sign}(E)}{k^2-k'^2+i\text{ sign}(E)\text{ }0^+}\nonumber\\
=\text{sign}(E)\frac{k}{2\pi\hbar v_F}\log |kR|-\frac{ik}{4\hbar v_F},
\end{align}
The integral is cut-off at momenta $k^{\prime}
\sim R^{-1}$, where $R$ is
of the order of the actual spatial range of the scatterer potential. 

\section{Solution of the Boltzmann Equation}\label{appAB}

In this appendix, in order to make the article self-contained, we review the solution of the linearized Boltzmann equation (BTE) obtained in  Ref.~\onlinecite{ferreira2014extrinsic}.
For an external DC electric field, the linearized BTE
takes the form:\cite{ziman1972principles}
\begin{equation}\label{eqAB1}
(-e) {\bf E} \cdot {\bf v_k} \text{ }\partial_{\epsilon} n^0(\epsilon_{\bf k}) =-\partial_{t} n_{\sigma} ({\bf k})|_{\mathrm{coll}},
\end{equation}
where  ${\bf E}=|\mathbf{E}|\hat{\mathbf{x}} $ is the applied electric field, $(-e)$ the electron charge, $n^0(\epsilon) = \left(e^{\left(\epsilon-\mu \right)/k_B T} + 1 \right)^{-1}$ is the equilibrium Fermi-Dirac distribution,  and 
\begin{align}
{\bf v}_{\bf k}= \zeta v_F (\cos\phi({\bf k}), \sin\phi({\bf k}))
\end{align}
is the carrier velocity in graphene with the angle $\phi({\bf k}) = \arctan k_y/k_x$ (not to be confused with with the free Hamiltonian eigenstate $| \phi_{\mathbf{k},\sigma} \rangle$); $\zeta=\pm 1$ is the band index ($+1$ for electrons, $-1$ for holes) and
$n_\sigma({\bf k})$ is the distribution function for electrons with spin projection $\sigma=\uparrow / \downarrow$ and Bloch wave-vector ${\bf k}$. The spin quantization axis is chosen to be the axis perpendicular to the graphene plane, which we take to be the z-axis.
The term $\partial_{t} n_{\sigma}({\bf k})|_{\mathrm{coll}}$ denotes the collision 
integral,\cite{ferreira2014extrinsic}
\begin{equation}
- \partial_{t} n_{\sigma}({\bf k}) |_{\mathrm{coll}}= \sum_{{\bf p},{\sigma'=\uparrow,\downarrow}} [n_\sigma({\bf k})-n_{\sigma'}({\bf p})]W_{\sigma' \sigma} ({\bf p, k}).
\end{equation}
$W_{\sigma' \sigma} ({\bf p, k})$ is the  scattering rate\cite{KohnLuttingerBTE} from state $\left({\bf k},\sigma\right)$ to $\left({\bf p},\sigma'\right)$ due to the presence of impurities:
\begin{align}
W_{\sigma' \sigma} ({\bf p, k}) = \frac{2\pi n_{imp}}{\hbar} 
| \langle \phi_{\mathbf{p} \sigma'} | T (\epsilon_{\bf p})| \phi_{\mathbf{k} \sigma} \rangle |^2 \delta\left( \epsilon_{\bf k} -\epsilon_{\bf p} \right).\notag \\
\end{align}
Here $T(\epsilon)$ is the T-matrix that has been explicitly obtained in Appendix \ref{appA} and $n_{imp}$ is the density of impurities. The linearized BTE can be solved exactly by the following ansatz for $\delta n_\sigma({\bf k}) = n_\sigma({\bf k})  - n^0(\epsilon_{\bf k})$:
\begin{equation}
\delta n_\sigma({\bf k})= \zeta v_F\left [ A_\sigma (k) \cos\phi({\bf k})+B_\sigma(k)  \sin\phi({\bf k}) \right].
\end{equation}
Introducing this ansatz in~\eqref{eqAB1} and 
setting $\phi({\bf k}) = 0$ and $\phi({\bf k}) = \frac{\pi}{2}$ for longitudinal and transverse response respectively, we obtain the following system of algebraic equations for $A_\sigma$ and $B_\sigma$:
\begin{align}
\sum_{\sigma'= \uparrow,\downarrow}A_{\sigma'}\Gamma^C_{\sigma'\sigma}+B_{\sigma'}\Gamma^S_{\sigma'\sigma}-A_\sigma\Gamma^I_{\sigma'\sigma} &= -X, \label{eq:ansatz1}\\
\sum_{\sigma'=\uparrow,\downarrow}B_{\sigma'}\Gamma^C_{\sigma'\sigma}-A_{\sigma'}\Gamma^S_{\sigma'\sigma}-A_\sigma\Gamma^I_{\sigma'\sigma} &= 0,\label{eq:ansatz2}
\end{align}
where $X \equiv -e|{\bf E}| \left( \frac{\partial n^0(\epsilon_{\bf k})}{\partial \epsilon}\right)$ and the coefficients $\Gamma$ are defined as
\begin{align}
\label{eq:gamma1}
\Gamma^I_{\sigma'\sigma} & = \int \frac{d^2{\bf p}}{(2\pi)^2}W_{\sigma' \sigma} ({\bf p, k}), \\
\label{eq:gamma2}
\Gamma^C_{\sigma'\sigma} & = \int \frac{d^2{\bf p}}{(2\pi)^2}\cos[\phi({\bf p})-\phi({\bf k})]W_{\sigma' \sigma} ({\bf p, k}), \\
\label{eq:gamma3}
\Gamma^S_{\sigma'\sigma} & = \int \frac{d^2{\bf p}}{(2\pi)^2}\sin[\phi({\bf p})-\phi({\bf k})]W_{\sigma' \sigma} ({\bf p, k}).
\end{align}
where $\phi({\bf p})-\phi({\bf k}) = \theta$ is the scattering angle. 

Note that time-reversal symmetry imposes several constraints on Eq.~\eqref{eq:ansatz1} and Eq.~\eqref{eq:ansatz2}. In particular,  it requires that $\Gamma^{I}_{\sigma'\sigma}=\Gamma^{I}_{\bar{\sigma}'\bar{\sigma}}$,  $\Gamma^{C}_{\sigma'\sigma}=\Gamma^{C}_{\bar{\sigma}'\bar{\sigma}}$, $\Gamma^S_{\sigma'\sigma}=-\Gamma^S_{\bar{\sigma}'\bar{\sigma}}$, $A_\sigma=A_{\bar{\sigma}}$, and $B_\sigma=-B_{\bar{\sigma}}$, where we further denote $\bar{\sigma }$ as the opposite spin of $\sigma$ to make the notation more compact. These relations are used to simplify the above system of equations.
Using these coefficients $\Gamma$ , Eq.~\eqref{eq:ansatz1} and Eq.~\eqref{eq:ansatz2} take the following form in terms of four relaxation times,
\begin{align}
&A_\sigma=A_{\bar{\sigma}}=\frac{\tau_{sk}\tau_{sk}^* \tau_{tr}}{\tau_{sk}\tau_{sk}^* +\tau_{tr}\tau_{tr}^*}X,  \\
&B_\sigma=-B_{\bar{\sigma}}=-\frac{\tau_{sk}\tau_{tr}^* \tau_{tr}}{\tau_{sk}\tau_{sk}^* +\tau_{tr}\tau_{tr}^*}X.
\end{align}
The four relaxation times, $\tau_{tr}$, $\tau_{tr}^{*}$, $\tau_{sk}$, and $\tau_{sk}^*$ are defined as follows:
\begin{align}
& \frac{1}{\tau_{tr}}=\Gamma^I_{\sigma\sigma}- \Gamma^C_{\sigma\sigma}+ \Gamma^I_{\sigma\bar{\sigma}}- \Gamma^C_{\sigma\bar{\sigma}}, \\
& \frac{1}{\tau_{tr}^*}=\Gamma^I_{\sigma\sigma}- \Gamma^C_{\sigma\sigma}+ \Gamma^I_{\sigma\bar{\sigma}}+ \Gamma^C_{\sigma\bar{\sigma}}, \\
& \frac{1}{\tau_{sk}}=\Gamma^S_{\sigma\sigma}+\Gamma^S_{\sigma\bar{\sigma}}, \\
& \frac{1}{\tau_{sk}^*}=\Gamma^S_{\sigma\sigma}-\Gamma^S_{\sigma\bar{\sigma}}.
\end{align}
The physical meaning of the relaxation times becomes more transparent if we use Eq.~\eqref{eq:Tmatrix_scat_amp} and Eq.~\eqref{eq:elastic} to express the scattering rate $W_{\sigma' ,\sigma}(\mathbf{p}  ,\mathbf{k} )$ in terms of the scattering amplitude:
\begin{equation}
W_{\sigma' ,\sigma}(\mathbf{p}  ,\mathbf{k} )=\frac{4\pi^2\hbar v_{F}^2 n_{imp}}{k}|f_{\sigma',\sigma}(\theta)|^2  \delta\left( \epsilon_{\bf k} -\epsilon_{\bf p} \right),
\end{equation}
where $\theta =\arccos\frac{\bf k \cdot p}{ k^2} $ is the scattering angle. Hence,
\begin{align}
\frac{1}{\tau_{tr}}&= n_{imp}v_F\sum_{\sigma'=\uparrow, \downarrow} \int  (1-\text{cos} \theta) 
\Big | f_{\sigma' \sigma}(\theta) \Big|^{2} d\theta \nonumber\\
&\equiv n_{imp} v_F \Sigma_{tr}, \\
\frac{1}{\tau^*_{tr}}&= n_{imp}v_F\sum_{\sigma'=\uparrow, \downarrow} \int  (1-\eta_{\sigma\sigma'}\text{cos} \theta) 
\Big | f_{\sigma' \sigma}(\theta) \Big|^{2} d\theta \nonumber\\
&\equiv n_{imp} v_F \Sigma^*_{tr}, \\
\frac{1}{\tau_{sk}}&= n_{imp}v_F\sum_{\sigma'=\uparrow, \downarrow} \int  \text{sin} \theta 
\Big | f_{\sigma' \sigma}(\theta) \Big|^{2} d\theta \nonumber\\
&\equiv n_{imp} v_F \Sigma_{sk}, \\
\frac{1}{\tau^*_{sk}}&= n_{imp}v_F\sum_{\sigma'=\uparrow, \downarrow} \int  \eta_{\sigma\sigma'}\text{sin} \theta 
\Big | f_{\sigma' \sigma}(\theta) \Big|^{2} d\theta \nonumber\\
&\equiv n_{imp} v_F \Sigma^*_{sk},
\end{align}
where $\Sigma_{tr}, \Sigma^*_{tr}, \Sigma_{sk}, \Sigma^*_{sk}$ are the single-scatterer scattering cross sections corresponding to the four different scattering rates, and
\begin{align}
\eta_{\sigma\sigma'}=
\begin{cases}
1&\text{if }\sigma'=\sigma,\\
-1&\text{if }\sigma'=\bar{\sigma}.
\end{cases}
\end{align}

Having solved the linearized BTE,  we are redady to compute the longitudinal current $j_{tr}$ (assumed to be along the $x$-axis) and spin Hall current $j_{sH}$ (assumed to be along the $y$-axis):
\begin{align}
j_{tr}&=-e\int \frac{d^2{\bf k'}}{(2\pi)^2}[n_\sigma({\bf k})+n_{\bar{\sigma}}({\bf k})] \zeta v_F \cos\phi({\bf k}) \nonumber\\
&=-\frac{e \zeta}{2\pi \hbar^2}\int  d\epsilon  \text{ } \epsilon \text{ }A_\sigma X\nonumber\\
&=-\frac{e}{2 \pi \hbar^2} \int d\epsilon | \epsilon | \frac{\tau_{sk}\tau_{sk}^* \tau_{tr}}{\tau_{sk}\tau_{sk}^* +\tau_{tr}\tau_{tr}^*} \left[-e|{\bf E}| \frac{\partial n^0(\epsilon_{\bf k})}{\partial \epsilon}\right] \nonumber\\
&=\frac{e^2}{h} \int d\epsilon\frac{| \epsilon |}{\hbar} \frac{\partial n^0(\epsilon_{\bf k})}{\partial \epsilon} \frac{\tau_{sk}\tau_{sk}^* \tau_{tr}}{\tau_{sk}\tau_{sk}^* +\tau_{tr}\tau_{tr}^*}|{\bf E}| \equiv \sigma_{tr}|{\bf E}|, \\
j_{sH}&=-e\int \frac{d^2{\bf k'}}{(2\pi)^2}[n_\sigma({\bf k})-n_{\bar{\sigma}}({\bf k})] \zeta v_F \sin \phi({\bf k}) \nonumber\\
&=-\frac{e \zeta}{2\pi \hbar^2} \int  d\epsilon \text{ }\epsilon \text{ }B_\sigma({\bf k})X\nonumber\\
&=\frac{e}{2 \pi \hbar^2} \int d\epsilon | \epsilon | \frac{\tau_{sk}\tau_{tr}^* \tau_{tr}}{\tau_{sk}\tau_{sk}^* +\tau_{tr}\tau_{tr}^*} \left[-e|{\bf E}| \frac{\partial n^0(\epsilon_{\bf k})}{\partial \epsilon}\right] \nonumber\\
&=-\frac{e^2}{h} \int d\epsilon\frac{| \epsilon |}{\hbar}  \frac{\partial n^0(\epsilon_{\bf k})}{\partial \epsilon} \frac{\tau_{sk}\tau_{tr}^* \tau_{tr}}{\tau_{sk}\tau_{sk}^* +\tau_{tr}\tau_{tr}^*}|{\bf E}|\notag \\
 &= \sigma_{sH}|{\bf E}|.
\end{align}
From the above expressions,  we can read off the expressions for the longitudinal transport and spin Hall conductivities, Eqs.~\eqref{eq:6} and Eq.~\eqref{eq:7}, respectively. The spin Hall angle measures the fraction of the charge current transformed into spin current, i.e.
\begin{equation}
\gamma	= \frac{j_{sH}}{j_{tr}} = 	\frac{\sigma_{sH}}{\sigma_{tr}}.
\end{equation}

In the zero temperature limit, the spin hall angle can be further simplified 
to yield the following expression:
\begin{equation}
\gamma=\frac{\sigma_{sH}}{\sigma_{tr}}=-\frac{\tau^*_{tr}}{\tau^*_{sk}}=
-\frac{\Sigma_{sk}^*}{\Sigma_{tr}^*}.
\end{equation}
\section{Circular  Scatterer}\label{appC}

In this appendix, we consider the extreme case for which
the couplings in the Rashba potential have opposite sign, that is, $\Delta_2/\Delta_1 = -1$, which has been termed `non-standard' Rashba in the main text. We show that, in this case, under the assumption of a circular (i.e. pill-box shaped) scatterer,
a solution of the scattering problem in terms of partial wave
waves is still possible. 

 However, what makes `non-standard' Rashba different for the `standard' Rashba SOC (i.e. $\Delta_2/\Delta_1 = 1$) is that the conserved 
quantity is not longer the total angular 
momentum projection along the $z$-axis, $J_z = l_z \pm \sigma_z/2 + s_z/2$ (where $\pm$ applies
to ${\bf K}_{\pm}$ respectively), but
$M_z = l_z \pm \sigma_z/2 - s_z/2$, with 
$l_z = x p_y - yp_y$. 

Below we consider in parallel the two 
cases and  study the scattering solutions for the following 
potentials:
\begin{align}
&V_{SR}(r)  = \left[ \texttt{v}_{0}+\Delta_{SR}(\tau_{z}\sigma_{x}s_{y}-\sigma_{y}s_{x})\right]\Theta(R-r), \\
&V_{NR}(r)  =\left[\texttt{v}_{0}+\Delta_{NR}(\tau_{z}\sigma_{x}s_{y}+\sigma_{y}s_{x})\right]\Theta(R-r),
\end{align}
where $R$ is the radius of the circular scatterer potential. Note that we have also considered a scalar potential $\text{v}_{0}$ in addition to the standard Rashba/Non-standard Rashba type spin-orbit coupling potential.
\subsection{Non-standard Rashba (NR) case}\label{app:nsr}

As mentioned in Sec.~\ref{sec:level2}, the conserved quantity that plays the role of angular momentum when the 
Rashba coupling is of the non-standard kind is $M_{z}= l_z \pm \sigma_z/2 - s_z/2$. Therefore, the scattering waves can be expanded the basis of eigenstates of $M_z$,

\begin{align}\label{app:5}
\psi_m(r,\theta) \sim & 
\left(\begin{array}{c}
A_m(r)e^{im\theta} \\
B_{m+1}(r)e^{i(m+1)\theta} \\
\end{array}\right)\eta_{\uparrow} \nonumber \\
+&
\left(\begin{array}{c}
C_{m-1}(r)e^{i(m-1)\theta} \\
D_m(r)e^{im\theta} \\
\end{array}\right)\eta_{\downarrow}
\end{align} 
This wavefunction is an eigenstate of $M_{z}$ where $M_{z}\psi_m(r,\theta)=m\hbar \psi_m(r,\theta)$. $A,B,C,D$ are the unknowns to be determined by solving the Schr\"odinger equation. The real space angle $\theta=\arctan \frac{y}{x}$ here should not be confused with the scattering angle defined in the previous appendix.  

For $r>R$, the Hamiltonian reduces to the kinetic term $H_0 = v_F(\pm \sigma_x p_x + \sigma_y p_y)$. In this region, the wavefunction can be expanded as follows:
\begin{widetext}
\begin{equation} \label{eq:outgoing_wf_NR}
\psi^>_m(r,\theta)=
\left(\begin{array}{c}
J_m(kr)e^{im\theta} \\
iJ_{m+1}(kr)e^{i(m+1)\theta}
\end{array}\right) \eta_{\uparrow}
+
t _{ m }^{ \uparrow\uparrow}\left(\begin{array}{c}
H_m(kr)e^{im\theta} \\
iH_{m+1}(kr)e^{i(m+1)\theta}
\end{array}\right)\eta_{\uparrow}
+
t _{ m-1 }^{ \uparrow\downarrow}\left(\begin{array}{c}
H_{m-1}(kr)e^{i(m-1)\theta} \\
iH_{m}(kr)e^{im\theta}
\end{array}\right)\eta_{\downarrow},
\end{equation}
\end{widetext}
where we have assumed that the incident electron is from the conduction band with spinor  $\eta_{\uparrow}$. $J_{m}(kr)$ is the Bessel function of the first kind and $H_{m}(kr)$ is the Hankel function of the first kind.  
$t_{m}^{\uparrow \uparrow}$ ($t_{m}^{\uparrow \downarrow}$) is related to the non-spin flip (spin flip) partial wave scattering amplitude, which must  if there is no scattering potential.

For $r<R$, we have to solve for the Dirac equation
\begin{equation}
\label{app:sch}
  (H_0+V_{NR}) \psi_{m}^<( r,\theta ) = E(k)\psi^<_m(r,\theta).
\end{equation}
The eigenvalue $E(k)$ is obtained by diagonalizing the Sch\"ordinger equation in momentum space,
\begin{equation}
E(k)=\texttt{v}_0+\xi\Delta_{NR}+\lambda\sqrt{\Delta^2_{NR}+k^2}.
\end{equation}
$\lambda=\pm 1$ and $\xi=\pm 1$ are two band indices. Only elastic scatterings are considered, therefore the scattered wavefunction must also be in the conduction band, i.e. $\lambda=+1$. Substituting Eq. (\ref{app:5}) into Eq. (\ref{app:sch}), we have to solve for a system of differential equations (in the basis of $\{A\uparrow,B\uparrow,A\downarrow,B\downarrow\}$):
\begin{widetext}
\begin{equation}
\left(\begin{array}{cccc}
-\epsilon_\xi & -i(\frac{d}{dr}+\frac{1}{r}(m+1)) & 0 & -2i\Delta_{NR} \\
i(\frac{d}{dr}+\frac{1}{r}m) & -\epsilon_\xi & 0 & 0 \\
0 & 0 & -\epsilon_\xi & -i(\frac{d}{dr}+\frac{1}{r}m) \\
2i\Delta_{NR} & 0 & i(-\frac{d}{dr}+\frac{1}{r}(m-1)) & -\epsilon_\xi \\
\end{array}\right)
\left(\begin{array}{c}
A_m(r) \\
B_{m+1}(r) \\
C_{m-1}(r) \\
D_m(r) \\
\end{array}\right)=0.
\end{equation}
\end{widetext}

Here $\epsilon_\xi  \equiv  E(k)-$v$_0 $. Let $A_m(r)=J_m(qr)$, where $q$ is some wave number to be determined, the system of differential equations can be solved as
\begin{align}
\psi^<_m(r,\theta)=&
\sum_{\xi=\pm1}
\alpha_\xi  \bigg[
\left(\begin{array}{c}
J_m(q_\xi r)e^{im\theta} \\
i\frac{q_\xi}{\epsilon_\xi}J_{m+1}(q_\xi r)e^{i(m+1)\theta} \\
\end{array}\right)\eta_{\uparrow} \nonumber \\
&+
\left(\begin{array}{c}
\xi \frac{q_\xi}{\epsilon_\xi}J_{m-1}(q_\xi r)e^{i(m-1)\theta} \\
i \xi  J_m(q_\xi r)e^{im\theta} \\
\end{array}\right) \eta_{\downarrow} \bigg],
\end{align}
where $q_\xi=\sqrt{\epsilon_\xi^2-2\xi\epsilon_\xi\Delta_{NR}}$ and $\alpha_{\xi}$ are some linear combination weights to be determined from boundary conditions.
By equating the wavefunction at the boundary of the circular potential, $\psi^<_m(R,\theta)=\psi^>_m(R,\theta)$, we can derive the partial wave scattering amplitudes, $t_{m}^{\uparrow \downarrow}$ and $t_{m}^{\uparrow \uparrow}$.

Finally, we shall establish the relationship between partial wave amplitudes $t_{m}$ defined in this section and the scattering amplitude $f(\mathbf{p} , \sigma' ; \mathbf{k},\sigma )$ defined in App. \ref{appA}. 
The total scattered wavefunction $\psi(r,\theta)$ is obtained by taking linear combination over all the partial waves, $\psi(r,\theta)=\sum_{m} c_{m} \psi^{>}_{m}(r,\theta) $.  Setting $c_{m} = i^{m}$  and suming Eq.~\eqref{eq:outgoing_wf_NR} over $m$ yields:

\begin{widetext}
\begin{align} \label{eq:scatter_wave}
\psi(r,\theta) &= \sum_{m=-\infty}^\infty i^{m} \left(
\left(\begin{array}{c}
J_m(kr)e^{im\theta} \\
i J_{m+1}(kr)e^{i(m+1)\theta}
\end{array}\right) \eta_{\uparrow}
+
t_{ m }^{ \uparrow\uparrow}\left(\begin{array}{c}
 H_m(kr)e^{im\theta} \\
i H_{m+1}(kr)e^{i(m+1)\theta}
\end{array}\right)\eta_{\uparrow}
+
t_{ m-1 }^{ \uparrow\downarrow}\left(\begin{array}{c}
H_{m-1}(kr)e^{i(m-1)\theta} \\
i H_{m}(kr)e^{im\theta}
\end{array}\right)\eta_{\downarrow}  \right)  \nonumber \\
& =  
e^{ikr\cos\theta}\left(\begin{array}{c}
1 \\
1
\end{array}\right)  \eta_{\uparrow} + \sum_{m=-\infty}^\infty i^{m} \left(
t_{ m }^{ \uparrow\uparrow}\left(\begin{array}{c}
 H_m(kr)e^{im\theta} \\
i H_{m+1}(kr)e^{i(m+1)\theta}
\end{array}\right)\eta_{\uparrow}
+
t_{ m-1 }^{ \uparrow\downarrow}\left(\begin{array}{c}
H_{m-1}(kr)e^{i(m-1)\theta} \\
i H_{m}(kr)e^{im\theta}
\end{array}\right)\eta_{\downarrow} \right) \nonumber \\
&\approx
e^{ikr\cos\theta}\left(\begin{array}{c}
1 \\
1
\end{array}\right)  \eta_{\uparrow} 
+ \left(f_{ \uparrow\uparrow}(\theta) \eta_{\uparrow}
+ f_{ \uparrow\downarrow}(\theta) \eta_{\downarrow} \right)
\left(\begin{array}{c}
1 \\
e^{i \theta}
\end{array}\right) \frac{e^{ikr}}{\sqrt{r}} \nonumber \\
& =
\psi_{in} + \psi_{sc}.
\end{align}
\end{widetext}
In the third line, we used the asymptotic form ($kr\gg1$) of Hankel function, $H_{m}(kr) \approx \sqrt{\frac{2}{\pi k r}}e^{ikr- m\pi/2 - \pi/4}$. We have also introduced the non spin flip (spin flip) scattering amplitude $f_{\uparrow\uparrow}(\theta)$ ($f_{\uparrow\downarrow}(\theta)$) which will be shown to be identical to the one defined in Eq.~\eqref{eq:Tmatrix_scat_amp},
\begin{equation}\label{eq:f_vs_partial_wave1}
f_{\uparrow \uparrow}(\theta) =\sum_{m=-\infty}^{\infty} \sqrt{\frac{2}{\pi k}} e^{-i\left( \frac{\pi}{4}-\frac{m \pi}{2}\right)}t_m^{\uparrow \uparrow} e^{i m\theta},
\end{equation}
\begin{equation}\label{eq:f_vs_partial_wave2}
f_{\uparrow \downarrow}(\theta) =\sum_{m=-\infty}^{\infty} i \sqrt{\frac{2}{\pi k}} e^{-i\left( \frac{\pi}{4}-\frac{m \pi}{2}\right)}t_m^{\uparrow \downarrow} e^{i m\theta}.
\end{equation}

The first term in Eq.~\eqref{eq:scatter_wave} is the incident wave propagating in the  $x$ direction and the second term is the scattered radial wave which contains spin up and spin down component. 

The current operator in direction $\mathbf{n}=\frac{\mathbf{r}}{r}$ is
\begin{equation}
\hat{J}=\left(\begin{array}{cc}
0 & e^{-i\theta} \\
e^{i\theta} & 0 
\end{array}\right),
\end{equation}
where $\theta= \arctan \frac{y}{x}$. Using Eq.~\eqref{eq:scatter_wave}, the incident current is given by $J_{in} =  \psi_{in}^{\dagger} \sigma_{x} \psi_{in}=2$ while the scattered current is given by
\begin{equation}
J_{sc} = \psi_{sc}^{\dagger}\hat{J}\psi_{sc}=\frac{2}{r} \left(  |f_{\uparrow\uparrow}(\theta) |^2 + | f_{\uparrow \downarrow}(\theta) |^2\right)
\end{equation}

Thus, the differential cross-section is given by
\begin{align}
\frac{d\sigma}{d\theta} &= \frac{ r J_{sc}}{J_{in}} \nonumber \\
&= | f_{\uparrow\uparrow}(\theta) |^2 + | f_{\uparrow\downarrow}(\theta)|^2.
\end{align}
The above form is clear that it is identical to Eq.~\eqref{eq:differential_cross_section} and the relationship between the partial wave amplitude and scattering amplitude is established in Eq.  \eqref{eq:f_vs_partial_wave1} and Eq.~\eqref{eq:f_vs_partial_wave2}.

\subsection{Standard Rashba (SR) case}
In the SR case, the conservation of angular momentum is restored. Therefore, the wavefunction in real space must be of the form
\begin{align}\label{RashbaPW}
\psi(r,\theta) \sim&
\left(\begin{array}{c}
A_n(r)e^{in\theta} \\
B_{n+1}(r)e^{i(n+1)\theta}
\end{array}\right) \eta_{\uparrow} \nonumber \\
+&
\left(\begin{array}{c}
C_{n+1}(r)e^{i(n+1)\theta} \\
D_{n+2}(r)e^{i(n+2)\theta} \\
\end{array}\right)\eta_{\downarrow},
\end{align}
where $A,B,C,D$ are the unknowns to be determined from Schr\"odinger equation. Following similar methods to solve for the NR case, we expand the wavefunction for $r>R$ as follows,
\begin{widetext}
\begin{equation}
\psi^>_n(r,\theta)=
\left(\begin{array}{c}
J_n(kr)e^{in\theta} \\
iJ_{n+1}(kr)e^{i(n+1)\theta}
\end{array}\right)\eta_{\uparrow}
+
t_{ n }^{ \uparrow\uparrow}\left(\begin{array}{c}
H_n(kr)e^{in\theta} \\
iH_{n+1}(kr)e^{i(n+1)\theta}
\end{array}\right)\eta_{\uparrow}
+
t_{ n+1 }^{ \uparrow\downarrow}\left(\begin{array}{c}
H_{n+1}(kr)e^{i(n+1)\theta} \\
iH_{n+2}(kr)e^{i(n+2)\theta}
\end{array}\right)\eta_{\downarrow},
\end{equation}
\end{widetext}
where we have assumed that the incident electron is from the conduction band with spinor  $\eta_{\uparrow}$. For $r<R$, we solve the corresponding Schr\"odinger equation:
\begin{equation}\label{RashbaScheq}
(H_0+V_{SR})\psi^<_n(r,\theta)=E(k)\psi^<_n(r,\theta),
\end{equation}
where the energy eigenvalue $E(k)$ is diagonalized to be
\begin{equation}
E(k)=\text{v}_0+\eta\Delta_{SR}+\lambda\sqrt{\Delta_{SR}^2+k^2}.
\end{equation}
Here $\lambda$ and $\xi$ are the two band indices. Note that the energy spectrum of SR and NR are the same. Following the same convention in NR case, we choose $\lambda=+1$.
To solve for the coefficients $A_n$, $B_n$, $C_n$ and $D_n$, we substitute Eq.~\eqref{RashbaPW} into Eq.~\eqref{RashbaScheq},
\begin{widetext}
\begin{equation}
\left(\begin{array}{cccc}
-\epsilon_\eta & -i(\frac{d}{dr}+\frac{1}{r}(n+1)) & 0 & 0 \\
i(-\frac{d}{dr}+\frac{1}{r}n) & -\epsilon_\eta & -2i\Delta_{SR} & 0 \\
0 & 2i\Delta_{SR} & -\epsilon_\eta & -i(\frac{d}{dr}+\frac{1}{r}(n+2)) \\
0 & 0 & i(-\frac{d}{dr}+\frac{1}{r}(n+1) & -\epsilon_\eta \\
\end{array}\right)
\left(\begin{array}{c}
A_n(r) \\
B_{n+1}(r) \\
C_{n+1}(r) \\
D_{n+2}(r) \\
\end{array}\right)=0.
\end{equation}
\end{widetext}
Here $\epsilon_\xi \equiv E(k)-\text{v}_0)$. Let $A_{n}=J_{n}(q_{\xi}r)$, the system of differential equations can be solved and the solution reads:
\begin{align}
\psi_n^{<}(r,\theta) =& \sum_{\xi=\pm1} \alpha_\xi \bigg[
\left(\begin{array}{c}
J_n(q_\xi r)e^{in\theta}  \\
 i\frac{\epsilon_\xi}{q_\xi} J_{n+1}(q_\xi r)e^{i(n+1)\theta} \\
\end{array}\right) \eta_{\uparrow} \notag  \\
+&
\left(\begin{array}{c}
-\xi \frac{\epsilon_\xi}{q_\xi} J_{n+1}(q_\xi r)e^{i(n+1)\theta} \\
-i \xi J_{n+2}(q_\xi r)e^{i(n+2)\theta}
\end{array}\right) \eta_{\downarrow} \bigg].
\end{align}
The wave number $q_\xi = \sqrt{\epsilon^2_\xi -2\xi \epsilon_\xi \Delta_{SR}} $ and $\alpha_{\xi}$ are linear combination weights to be determined from the boundary conditions.
Using the boundary condition at the boundary of circular potential $\psi_n^>(R,\theta)= \psi_n^<(R,\theta)$, the partial wave amplitudes (i.e. $t_{n}^{\uparrow\uparrow}$ and $t_{n}^{\uparrow\downarrow}$) can be determined, hence the scattering amplitudes (i.e. $f_{\uparrow\uparrow}(\theta)$ and $f_{\uparrow\downarrow}(\theta)$), see Eq.~\eqref{eq:scatter_wave}.

Using the solution of the Boltzmann equation described in Appendix~\ref{appAB}, the spin Hall angle $\gamma$  for both SR and NR case can be computed using the scattering amplitudes.  We have plotted the results in Fig.~\ref{SR and NR comparison} showing the $\text{v}_0$ dependence of $\gamma$ for both SR and NR cases.

\bibliography{MyBIB}

\end{document}